\documentclass[twocolumn,aps,prb,10pt,superscriptaddress,longbibliography]{revtex4-2}

\usepackage{amsmath}
\usepackage{bm}
\usepackage{booktabs}
\usepackage{graphicx}
\usepackage{mathtools}
\usepackage{multirow}
\usepackage{textcomp}
\usepackage[usenames,dvipsnames,svgnames]{xcolor}
\usepackage[english]{babel}
\usepackage{xurl}
\usepackage{hyperref}
\hypersetup{
  pdfnewwindow=true,
  colorlinks=true,
  linkcolor=Blue,
  citecolor=Blue,
  filecolor=Blue,
  urlcolor=Blue
}
\usepackage{siunitx}
\DeclareSIUnit\angstrom{\text{\AA}}
\DeclareSIUnit\atom{\text{atom}}

\newcommand{\DFT}{DFT}
\newcommand{\MLP}{MLP}
\newcommand{\NEP}{NEP}
\newcommand{\GPUMD}{GPUMD}
\newcommand{\EqVthree}{EqV3}

\newcommand{\Dzero}{D00}
\newcommand{\Dhundred}{D100}

\newcommand{\BAsGaN}{BAs/GaN}
\newcommand{\CuDiamond}{Cu--diamond}

\begin{document}
% \linenumbers

\title{Distilling an atomistic foundation model is all you need for large-scale molecular dynamics of semiconductor interfaces}
% Direct distillation of atomistic foundation models for thermal and mechanical simulations of heterogeneous interfaces

\author{Feiyu Zhou}
% \email{zhoufy20@stu.xjtu.edu.cn}
\thanks{These authors contributed equally to this work.}
\affiliation{Laboratory for multiscale mechanics and medical science, SV LAB, School of Aerospace, Xi'an Jiaotong University, Xi'an, Shaanxi, 710049, China}

\author{Wei Qiu}
% \email{weiqiu@tauex.tau.ac.il}
\thanks{These authors contributed equally to this work.}
\affiliation{Department of Physical Chemistry, School of Chemistry, The Raymond and Beverly Sackler Faculty of Exact Sciences and The Sackler Center for Computational Molecular and Materials Science, Tel Aviv University, Tel Aviv 6997801, Israel}

\author{Fei Shuang}
\email{shuangfei@imech.ac.cn}
\affiliation{State Key Laboratory of Nonlinear Mechanics, Institute of Mechanics, Chinese Academy of Sciences, Beijing 100190, China}

\author{Yilun Liu}
% \email{yilunliu@mail.xjtu.edu.cn}
\affiliation{Laboratory for multiscale mechanics and medical science, SV LAB, School of Aerospace, Xi'an Jiaotong University, Xi'an, Shaanxi, 710049, China}

\author{Shengying Yue}
\email{syyue@xjtu.edu.cn}
\affiliation{Laboratory for multiscale mechanics and medical science, SV LAB, School of Aerospace, Xi'an Jiaotong University, Xi'an, Shaanxi, 710049, China}

\author{Penghua Ying}
\email{penghua@xjtu.edu.cn}
\affiliation{Laboratory for multiscale mechanics and medical science, SV LAB, School of Aerospace, Xi'an Jiaotong University, Xi'an, Shaanxi, 710049, China}

\date{\today}

\begin{abstract}
\vspace{0.5cm}
\noindent\textbf{Abstract}
\noindent
State-of-the-art atomistic foundation models provide broadly transferable energy and force predictions, but their direct deployment in large-scale molecular dynamics remains computationally costly. Here we examine whether their knowledge can be distilled into compact potentials while retaining the physical responses of heterogeneous semiconductor interfaces. From 128 pretrained models spanning 22 model families, we select a fixed EquiformerV3 (\EqVthree{}) teacher and train atomic cluster expansion (ACE) and neuroevolution potential (\NEP{}) students without target-specific teacher fine-tuning, with both simulation speed and accessible system size improved by several orders of magnitude relative to the teacher model. Matched training configurations and architectures enable direct comparisons with students trained on density-functional-theory (\DFT{}) reference labels. For 79,920-atom \BAsGaN{} interfaces, direct distillation preserves termination-dependent thermal conductance and dominant spectral features, with conductance differing by at most 6.6\% from \DFT{}-trained counterparts. In 29,792-atom \CuDiamond{} specimens loaded at 500\,K, tensile strength and loading work differ by 2.8\% and 1.5\%, respectively. Exploratory SiC/a-C/SiC simulations retain the principal thermal bottleneck and the direction of the annealing response. Furthermore, a four-model committee retrieves 42--56\% of the highest-error teacher configurations with a 10\% selection budget and improves high-error force fitting, but does not consistently improve downstream properties. These findings demonstrate that direct distillation can preserve key interfacial thermal and mechanical responses in compact interatomic potentials, opening a path towards large-scale, high-throughput simulations of heterogeneous interfaces.
\end{abstract}

\maketitle

\noindent\textbf{INTRODUCTION}

\noindent Heterogeneous integration expands the functionality of electronic and photonic devices by combining materials with complementary properties. For example, single-crystalline nanomembranes enable electro-optic modulation and optical isolation on silicon photonic platforms~\cite{Meng2026HeterogeneousPhotonics}; similarly, boron arsenide (BAs) cooling substrates reduce hot-spot temperatures in gallium nitride (GaN) transistors~\cite{Kang2021GaNBAsIntegration}. Device performance also depends on interfacial heat transport and mechanical integrity. For instance, annealing-induced covalent bonding through a SiO$_2$ interlayer enhances thermal conductance in Ga$_2$O$_3$/SiC heterostructures~\cite{Shen2025Ga2O3SiC}. Interfacial bonding also influences adhesion and mechanical failure~\cite{Mortazavi2021InterfaceMechanics}. Relating these responses to atomic-scale bonding, vibrational coupling, and structural evolution remains challenging, particularly for buried interfaces. Atomistic simulations therefore provide a complementary means of investigating the microscopic mechanisms underlying interfacial thermal and mechanical responses.

The computational cost of density-functional-theory (\DFT{})-based molecular dynamics (MD) limits its ability to capture interfacial processes requiring large simulation cells and long trajectories. Machine learning potentials (\MLP{}s) approximate high-dimensional potential-energy surfaces by learning from first-principles reference data~\cite{Behler2007HDNNP,Bartok2010GAP}. These learned potentials can approach first-principles accuracy at substantially lower computational cost, extending the accessible length and time scales~\cite{Deringer2019-AM-review}. For example, \MLP{}-based simulations of \BAsGaN{} have related interfacial conductance to vibrational matching and atomic-scale interface structure~\cite{Wu2024GaNBAs,Yu2026GaNBAs}. Alloy and amorphous interlayers modify phonon transmission at GaN/SiC junctions~\cite{Zhang2026GaNSiCAmorphousInterlayer}, while orientation, strain and inserted layers influence thermal transport across Cu/diamond interfaces~\cite{Sun2025CuDiamond}. By tuning interfacial phonon modes, \MLP{}-based simulations have also predicted an ideal upper limit for thermal conductance across diamond/$c$BN semiconductor heterointerfaces~\cite{Wang2026DiamondCBNPhonons}. Beyond thermal transport, \MLP{}-based simulations have linked preparation history to tensile response in Cu/diamond~\cite{Zhou2026NEPCuCPreprint} and interfacial chemistry to fracture-plane selection in diamond/$c$BN heterostructures~\cite{Qiu2026DiamondCBNFracture}. The eleven-element SemiCID database has extended these capabilities through universal and task-specific potentials for semiconductor thermal transport, including multiple heterointerfaces and roughness-dependent transport in AlAs/GaAs superlattices~\cite{Hu2026SemiCID}. However, developing potentials for new material combinations and evolving interfacial structures can still require extensive \DFT{} reference data.

The growing availability of atomistic foundation models is accelerating computational materials discovery and simulation. Matbench Discovery provides standardized comparisons of diverse pretrained models for crystal-stability prediction~\cite{Riebesell2025MatbenchDiscovery}. Large electronic-structure datasets, including Materials Project Trajectory, Alexandria and Open Materials 2024, support atomistic foundation models with broad chemical and structural coverage~\cite{deng2023chgnet,Schmidt2024Alexandria,barrosluque2026omat24}. MatPES and MP-ALOE add off-equilibrium structures and higher-fidelity \DFT{} labels, while ELEMENTA broadens systematic sampling across composition and configuration space~\cite{Kaplan2025MatPES,Kuner2025MPALOE,Kairos2026ELEMENTA}. Models including ALIGNN-FF, MACE, CHGNet, DPA and Equiformer provide reusable energy and force predictions across different material systems~\cite{Choudhary2023ALIGNNFF,Batatia2025-MACE-JCP,deng2023chgnet,Zhang2024DPA2,liao2026equiformerv3}.

For large-scale interfacial MD, direct use of a pretrained model requires both accurate target-property predictions and affordable repeated force evaluation. Large graph-based models can impose substantial inference and memory costs~\cite{Kong2026ScalableFoundation}. Atomic cluster expansion (ACE) and neuroevolution potential (\NEP{}) frameworks support compact implementations for efficient dynamics through LAMMPS and \GPUMD{}~\cite{Drautz2019ACE,Fan2021NEP,LAMMPS,Xu2025-GPUMD}, and can underpin both universal pretrained models and system-specific potentials~\cite{lysogorskiy2026grace,liang2026nep89}. Published GPU examples reach approximately $2\times10^{5}$\,atom-steps\,s$^{-1}$ for an ACE potential in a 10,000-atom phase-change material on an NVIDIA A100~\cite{Zhou2025DeviceScaleACE} and $5\times10^{6}$\,atom-steps\,s$^{-1}$ for NEP89 in a twenty-element alloy on an NVIDIA RTX 4090~\cite{liang2026nep89}. Such throughput makes fast pretrained potentials attractive for direct use, but their accuracy for a particular interface property still requires validation~\cite{Morrow2023Validation}. Distillation offers a route to reuse predictions from an accurate, computationally demanding teacher through compact potentials suited to repeated production simulations.

Knowledge distillation transfers teacher predictions or learned representations into compact interatomic potentials. Approaches include fitting teacher-generated energies and forces~\cite{Morrow2022IndirectLearning,Gardner2025CrossArchitectureDistillation}, matching teacher energy Hessians~\cite{Amin2025HessianDistillation}, and aligning angular relations between latent representations~\cite{Lim2026RelationalDistillation}. Distillation has produced the lightweight universal SevenNet-Nano potential~\cite{oh2026sevennetnano} and application-specific students trained after teacher fine-tuning for target materials, including doped solid-state electrolytes~\cite{Wang2025PFD,Zhang2026MinimumAbInitio,Wu2026PFDElectrolytes}. Lian and Pasquarello integrated active learning with MACE foundation-model fine-tuning and subsequent distillation into a fast DeePMD potential for liquid water~\cite{Lian2026ActiveLearningDistillation}. LightPFP used a pretrained PFP teacher without target-specific fine-tuning to train MTP-based students for metal--liquid interfacial thermal transport and graphene nanoribbon fracture~\cite{Li2025LightPFP}. The GRACE study also compared GRACE-FS students trained on either teacher-relabeled or original \DFT{}-labeled HEA25S alloy configurations~\cite{lysogorskiy2026grace}.

Uncertainty estimation offers complementary tools for identifying predictions that may require additional reference calculations. In Au/water simulations, atom-resolved uncertainty analysis exposed localized interfacial failures despite improvements in global model accuracy~\cite{Bilbrey2026InterfaceUQ}. Other studies have developed shared-backbone multi-head MACE committees for force uncertainty~\cite{Beck2025MultiHeadUQ} and post hoc calibration of force-uncertainty estimates~\cite{Ho2026FlexibleCalibration}. Liu \textit{et al.} used heterogeneous pretrained-model disagreement to guide selective \DFT{} relabeling for W. For MoNbTaW, they trained ACE potentials using only foundation-model labels and validated their mechanical response~\cite{Liu2026HeterogeneousEnsemble}. Model accuracy and data heterogeneity affect uncertainty--error relationships~\cite{Shuang2026-ACE-UQ}, so error-ranking performance must be validated for each target system. These studies establish complementary routes to compact potentials, including direct distillation and uncertainty-guided \DFT{} correction.

Previous studies have established direct distillation, alternative training pathways and physical validation in selected systems~\cite{Li2025LightPFP,lysogorskiy2026grace}. However, agreement between directly distilled and matched \DFT{}-trained students remains insufficiently established across interfacial heat transport, nonlinear deformation and processing-induced disorder. Interfacial conductance probes collective vibrational energy transfer~\cite{Chen2022ITRReview}, whereas nonlinear loading and thermal processing generate evolving atomic environments. Low energy and force errors alone cannot establish agreement in these physical responses~\cite{Creed2026SixQuestions,Morrow2023Validation,fu2025esen,Chiang2025MLIPArena}. It is also unclear whether selectively correcting inaccurate teacher labels improves the corresponding physical predictions. We address these questions through controlled comparisons of labeling strategies, holding training configurations, student architecture and fitting protocol fixed within each material system.

Here we screen 128 pretrained atomistic foundation-model checkpoints spanning 22 model families and select EquiformerV3 (\EqVthree{}) as the teacher. We distill the teacher's energy and force predictions into compact ACE and \NEP{} potentials for production molecular dynamics, using the frozen teacher for offline labeling without target-specific fine-tuning. MD benchmarks show that a student potential can achieve nearly three orders of magnitude higher throughput than the teacher. We compare these students with counterparts trained on \DFT{} reference labels and with students trained after selective \DFT{} correction of teacher labels guided by a four-model committee.
We also evaluate frozen GRACE-FS-OAM and NEP89 as fast universal pretrained alternatives within the ACE and \NEP{} families, respectively~\cite{lysogorskiy2026grace,liang2026nep89}. GRACE-FS-OAM is used for \CuDiamond{} loading and NEP89 for \BAsGaN{} and SiC/a-C/SiC transport (a-C denotes amorphous carbon). These baselines test whether the investigated physical responses can already be reproduced through direct deployment without constructing system-specific students, thereby assessing the practical benefit of distillation. The physical tests begin with termination-dependent \BAsGaN{} heat flow and extend to the response of \CuDiamond{} as loading drives structural evolution. Exploratory SiC/a-C/SiC simulations then examine transport through disordered interlayers generated and annealed with each potential. We find that direct distillation retains several key interfacial responses, whereas committee-guided correction improves force fitting in high-error training regions without consistently improving downstream physical properties.

\begin{figure*}[tp]
    \centering
    \includegraphics[width=0.8\textwidth]{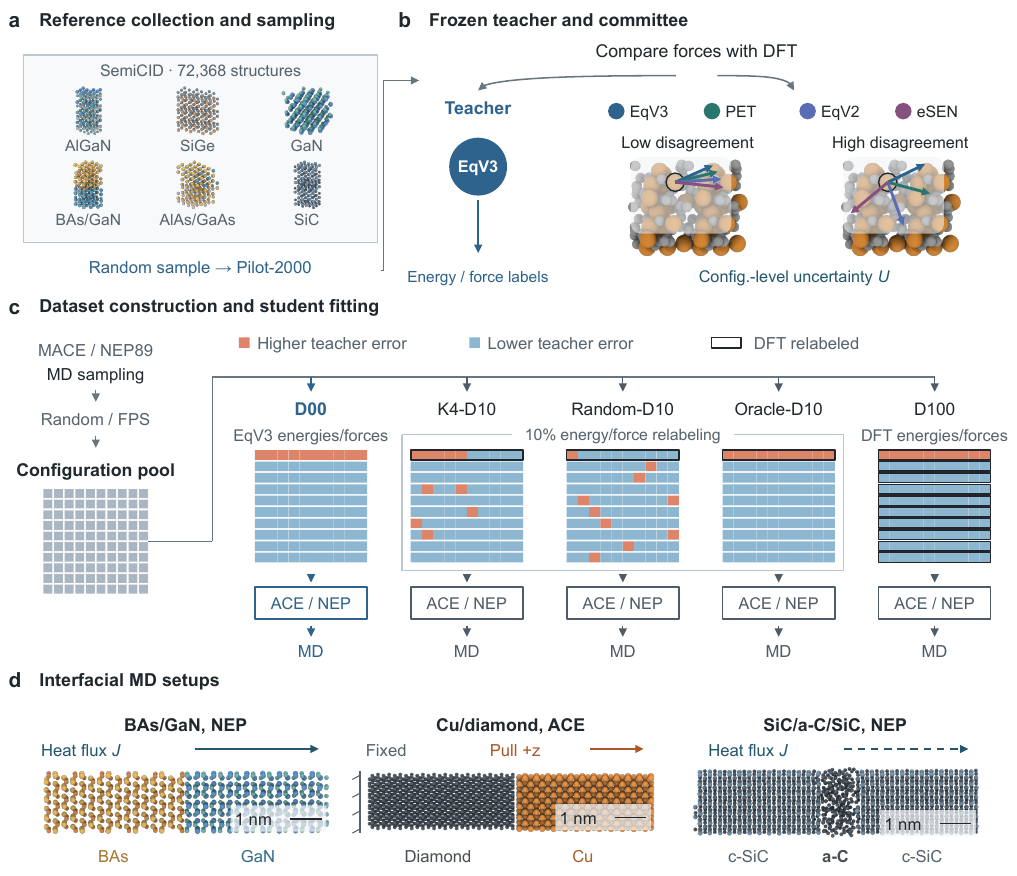}
    \caption{\textbf{Controlled comparisons of direct distillation and selective \DFT{} correction.} \textbf{a}, SemiCID structures and Pilot-2000. \textbf{b}, Frozen teacher and four-model committee. \textbf{c}, Additional sampling for SiC/carbon, training-label strategies and student fitting. \Dzero{} and \Dhundred{} use teacher and \DFT{} labels, respectively. The D10 routes replace a nominal $10\%$ of training configurations' energy and force labels using committee disagreement (K4-D10), random selection (Random-D10) or retrospectively known teacher errors (Oracle-D10). Outlines mark \DFT{} labels, and orange/blue indicate higher/lower teacher errors. Each schematic row represents $10\%$ of a target training set, arranged independently across strategies. \textbf{d}, \BAsGaN{} transport, \CuDiamond{} loading and SiC/a-C/SiC thermal processing and heat flow.}
    \label{fig:framework}
\end{figure*}

\vspace{0.5cm}
\noindent\textbf{RESULTS}

\vspace{0.2cm}
\noindent\textbf{Transferring pretrained atomic knowledge into compact potentials}

\noindent Knowledge distillation was developed for model compression, training compact students to reproduce the predictions of larger teachers~\cite{Bucila2006ModelCompression,Hinton2015Distillation}. In large language models, distillation has been used to transfer language generation and reasoning capabilities from larger teachers to smaller students~\cite{Gu2024MiniLLM,Guo2025DeepSeekR1}. Here, we applied this principle to atomistic modeling by fitting compact potentials to a pretrained foundation model's energy and force predictions for interfacial dynamics (Fig.~\ref{fig:framework}). The teacher remained frozen across target systems and supplied labels offline without target-specific fine-tuning, while compact students performed production MD.

\begin{figure}[tp]
    \centering
    \includegraphics[width=\columnwidth]{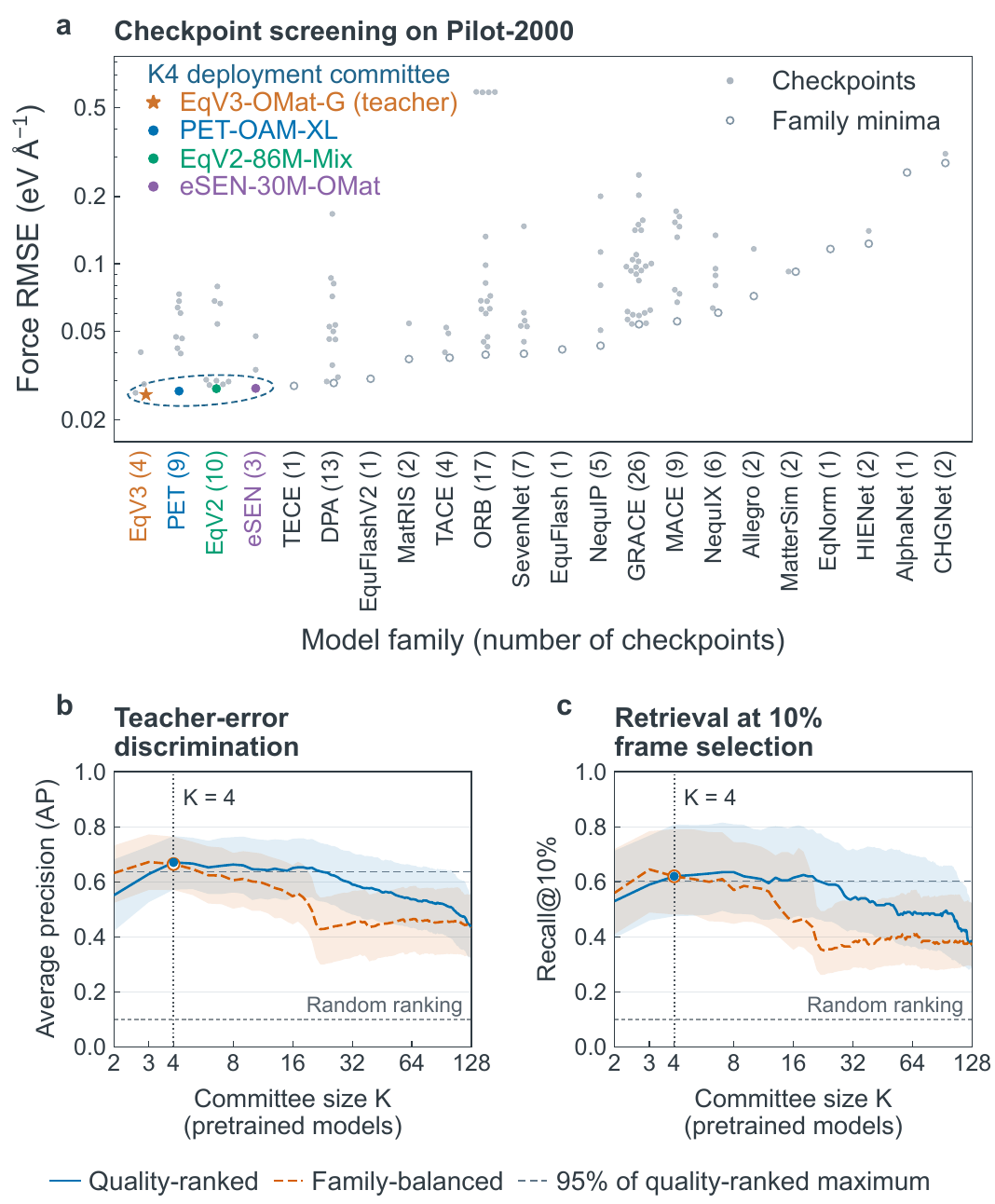}
    \caption{\textbf{Teacher and committee selection.} \textbf{a}, Force-component RMSE for 128 checkpoints in 22 families on Pilot-2000. Points denote checkpoints, open circles family minima, and colored markers the teacher and committee members. \textbf{b,c}, Out-of-fold AP and Recall@10\% versus committee size on the same development set. Guides mark $K=4$, random ranking and the $95\%$ performance-retention thresholds. Shading shows pointwise $95\%$ block-bootstrap intervals. Committee construction and validation are described in Methods.}
    \label{fig:pilot-families}
\end{figure}

Because the teacher supplied labels offline, force accuracy took priority over inference speed in its selection. We screened 128 released checkpoints from 22 model families using Pilot-2000, a development set of 2,000 \DFT{}-labeled configurations randomly sampled from SemiCID (Figs.~\ref{fig:framework}a and~\ref{fig:pilot-families}a). SemiCID covers eleven elements and a range of semiconductor materials and heterointerfaces~\cite{Hu2026SemiCID}. Pilot-2000 provided a common reference for teacher selection and committee development. Force errors varied both across families and among checkpoints within a family. EquiformerV3 OMat-G (\EqVthree{}) achieved the lowest force-component root-mean-square error (RMSE), $0.02597$\,eV\,\AA$^{-1}$, and was selected as the sole teacher~\cite{liao2026equiformerv3}.

\EqVthree{} is an $\mathrm{SE}(3)$-equivariant graph attention Transformer with approximately 30.3 million parameters, using local atomic environments to predict the total energy. Its equivariant representation keeps the predicted energy unchanged under rigid translations and rotations, while the corresponding forces rotate with the atomic structure. Atomic forces are computed as negative energy gradients with respect to atomic positions, providing consistent energy and force labels for distillation. The released OMat-G checkpoint had undergone direct pretraining followed by gradient fine-tuning on the OMat24 inorganic-materials dataset~\cite{barrosluque2026omat24}. We used this checkpoint without additional target-specific fine-tuning.

To examine the Pilot-selected teacher's target-domain accuracy, we retrospectively audited the same 128 checkpoints on the Cu--C and \BAsGaN{} training sets. The checkpoint force-error rankings on these two target datasets were correlated (Spearman $\rho=0.865$), and \EqVthree{} remained among the lower-error models in both domains (Supplementary Fig.~S1). The Pilot-selected checkpoint was retained without reselection. Model sources and checkpoint rankings are given in Supplementary Tables~S1 and S2, respectively.

The central test was whether compact students retained the physical responses of matched \DFT{}-trained counterparts. We compared directly distilled (\Dzero{}) and \DFT{}-trained (\Dhundred{}) students (Fig.~\ref{fig:framework}c). The suffixes specify the percentage of training configurations assigned \DFT{} energy and force labels. ACE-D00 and ACE-D100 described Cu--C, whereas NEP-D00 and NEP-D100 described \BAsGaN{} and SiC.

\vspace{0.3cm}
\noindent\textbf{Uncertainty-guided correction of teacher labels}

\noindent A low force-component RMSE can conceal severe local force errors. We therefore asked whether disagreement within a committee of pretrained models could identify the frozen \EqVthree{} teacher's most inaccurate labels for selective \DFT{} correction (Fig.~\ref{fig:framework}b)~\cite{Liu2026HeterogeneousEnsemble}. Each configuration's teacher error was defined as its largest atomic force-vector error against \DFT{}. \EqVthree{} alone supplied teacher labels, while the committee ranked configurations for correction. We combined each member's largest atomic force-vector deviation from the committee mean into a weighted configuration-level score $U$ (Methods). The score provided a relative risk ranking without assigning a calibrated error bound. Disagreement was positively associated with teacher force error on Pilot-2000 (Supplementary Fig.~S2).

Candidate committees were constructed from the 128 screened checkpoints using two rules. The quality-ranked rule added checkpoints in order of increasing force RMSE, whereas the family-balanced rule first added each family's lowest-error checkpoint in order of accuracy (Methods). Grouped five-fold evaluation on Pilot-2000 produced out-of-fold scores, pooled to assess retrieval of the highest-error decile using average precision (AP) and Recall@10\% (Fig.~\ref{fig:pilot-families}b,c). Recall@10\% measures the fraction of the highest-error decile recovered among the highest-scoring $10\%$ of configurations. AP summarizes how well the disagreement ranking prioritizes high-error configurations across selection thresholds. Higher values indicate better retrieval.

Along the quality-ranked curves, $K=4$ was the smallest committee size retaining at least $95\%$ of the maximum AP and maximum Recall@10\%. At this size, family balancing gave comparable retrieval with heterogeneous architectures from four families and was adopted for the final committee. This strategy achieved out-of-fold AP $0.665$ and Recall@10\% $0.620$. Selecting the highest-scoring $10\%$ of configurations therefore recovered $62.0\%$ of the teacher's highest-error decile, compared with $10\%$ expected from random selection. Because Pilot also determined committee size, these values describe development-set performance. Expanding the family-balanced committee to 22 families reduced AP and Recall@10\% to $0.428$ and $0.355$, supporting the compact committee under the tested construction rule. At $K=4$, the family-balanced AP also exceeded that of all 100 sampled random four-model committees and changed little across the three disagreement definitions $U$ (Supplementary Fig.~S3 and Supplementary Note~2).

Applying the selected family-balanced rule to the full Pilot set yielded \EqVthree{}-OMat-G, PET-OAM-XL, EqV2-86M-Mix and eSEN-30M-OMat (Supplementary Table~S3). We froze membership and weights before applying the committee to the target training sets used for student fitting. These comprised 950 Cu--C configurations from our previous study~\cite{Zhou2026NEPCuCPreprint}, 2,848 \BAsGaN{} configurations from SemiCID, and 1,752 exploratory SiC/carbon configurations from existing datasets and additional MD sampling (Methods). Selecting the highest-$U$ $10\%$ of each training set recovered $55.8\%$, $42.1\%$ and $46.6\%$ of the teacher's highest-error decile in these three systems, respectively (Supplementary Figs.~S4 and S5). The selected subsets were thus enriched in configurations with large teacher errors, although more than $40\%$ of the highest-error decile remained unselected in each system.

We compared five labeling strategies using identical training configurations, student architectures and fitting protocols within each system (Fig.~\ref{fig:framework}c). \Dzero{} used energy and force predictions from the frozen \EqVthree{} teacher, whereas \Dhundred{} used \DFT{} references throughout. The three D10 strategies replaced teacher energy and force labels with \DFT{} references for a nominal $10\%$ of the training configurations. K4-D10 selected the highest-$U$ configurations, Random-D10 selected uniformly at random, and Oracle-D10 selected configurations with the largest retrospectively known teacher errors. Oracle-D10 provided an error-informed selection control without defining an upper bound on student accuracy.

The \Dzero{}--\Dhundred{} comparison tested whether direct distillation retained the force accuracy and physical responses of matched \DFT{}-trained students. Comparing each D10 strategy with \Dzero{} assessed whether partial \DFT{} correction improved upon direct distillation. Within the D10 group, comparisons of K4-D10 with Random-D10 and Oracle-D10 evaluated committee-guided selection against random and error-informed controls. Before student fitting, K4-D10 reduced training-label force RMSE relative to the Random-D10 median by $16.7\%$, $1.2\%$ and $27.6\%$ for Cu--C, \BAsGaN{} and SiC, respectively (Supplementary Fig.~S6 and Supplementary Note~3). We next assessed the efficiency of the chosen student frameworks and compared the force accuracy obtained with each labeling strategy.

\vspace{0.3cm}
\noindent\textbf{Computational efficiency and force fidelity of student potentials}
\par\nobreak

\begin{figure}[tp]
    \centering
    \includegraphics[width=\columnwidth]{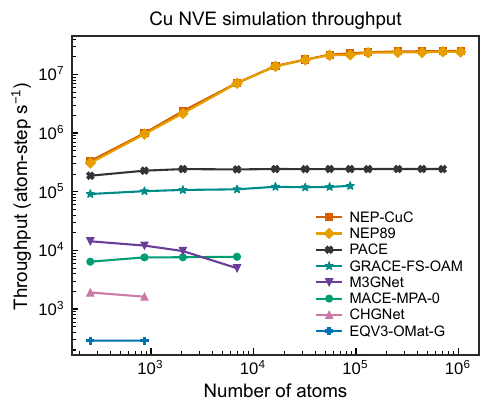}
    \caption{\textbf{Cu NVE simulation throughput.} Periodic FCC Cu on a single NVIDIA RTX 4090 D. Symbols show median throughput from at least three independent process launches; lines connect measured sizes. PACE denotes the Cu--C ACE-D00 student and EQV3-OMat-G the frozen teacher. The NEP-CuC potential and benchmark protocol are from our previous study~\cite{Zhou2026NEPCuCPreprint}.}
    \label{fig:cu-nve-throughput}
\end{figure}

\noindent We used ACE and \NEP{} as compact student potentials for efficient production MD through LAMMPS and \GPUMD{}, respectively~\cite{Drautz2019ACE,Lysogorskiy2021PACE,Fan2021NEP,Xu2025-GPUMD}. Cu NVE benchmarks compare representative ACE/\NEP{} implementations with the \EqVthree{} teacher and open-source pretrained potentials, including NEP89, GRACE-FS-OAM, MACE-MPA-0, CHGNet and M3GNet (Fig.~\ref{fig:cu-nve-throughput}). The compact ACE and \NEP{} implementations delivered orders of magnitude higher throughput than the teacher, providing the efficiency required for large-scale interfacial MD.

\begin{figure*}[tp]
    \centering
    \includegraphics[width=0.9\textwidth]{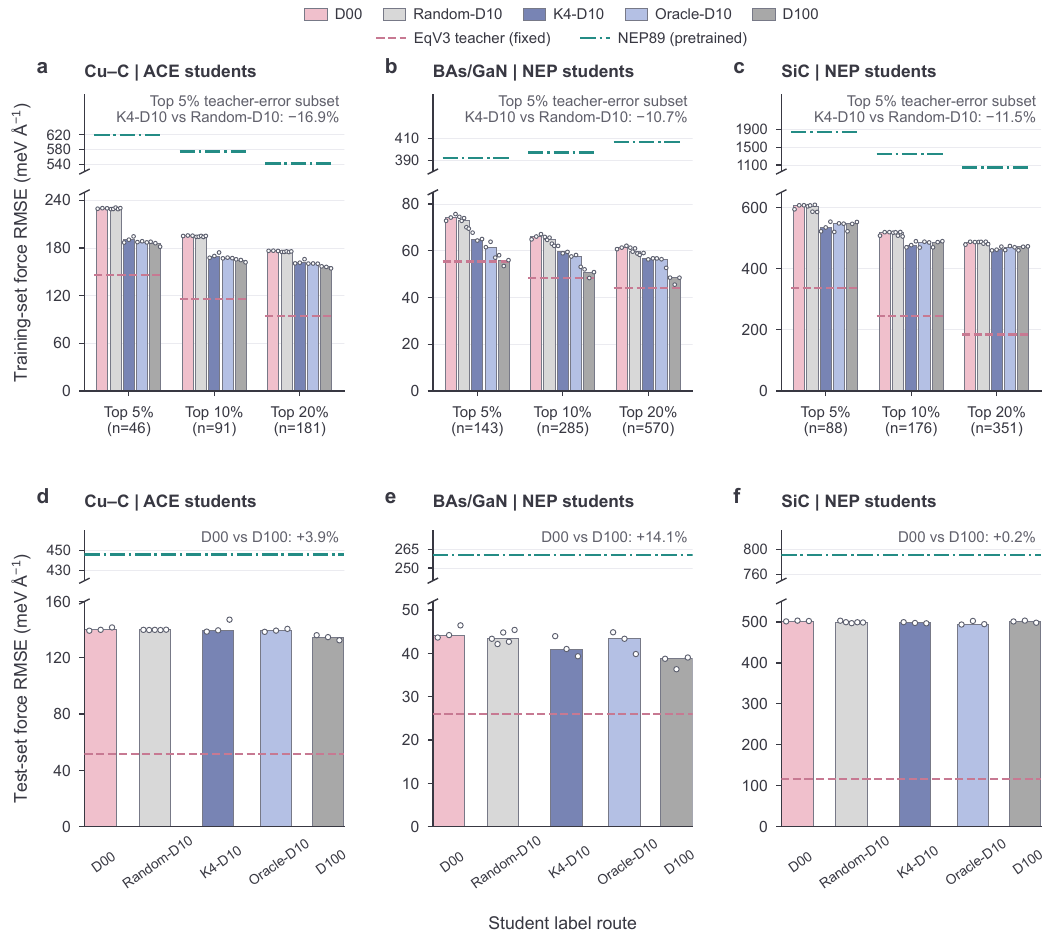}
    \caption{\textbf{Student force accuracy across labeling strategies.} \textbf{a--c}, Training-set force-component RMSE on subsets ranked retrospectively by teacher force error (top $5\%$, $10\%$ and $20\%$). \textbf{d--f}, Test-set RMSE on 50 Cu--C, 170 \BAsGaN{} and 195 SiC configurations. Cu--C panels use ACE potentials. \BAsGaN{} and SiC panels use \NEP{} potentials. Legend codes identify the labeling strategies. Bars show medians, with points representing three individual potentials per fixed strategy or five selection medians for Random-D10. Reference lines show frozen \EqVthree{} and NEP89 predictions. Broken axes omit empty ranges. All errors are relative to \DFT{}. Test-set provenance is given in Methods.}
    \label{fig:training-risk}
\end{figure*}

% Queue the remaining main figures for consecutive one-per-page placement.
\begin{figure*}[tp]
    \centering
    \includegraphics[width=0.8\textwidth]{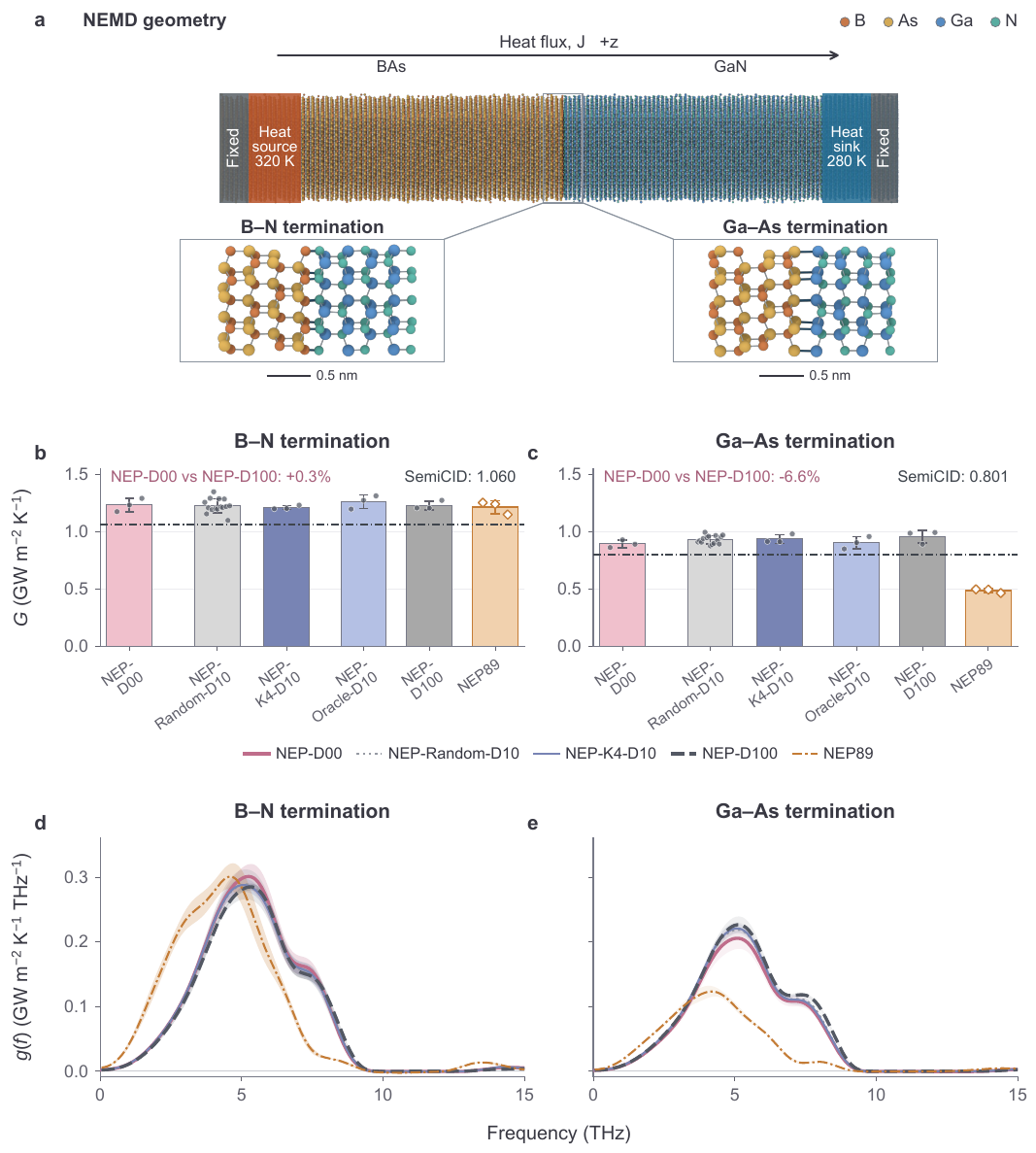}
    \caption{\textbf{\BAsGaN{} interfacial heat transport.} \textbf{a}, NEMD geometry and B--N/Ga--As terminations. Bonds indicate geometric neighbors. \textbf{b,c}, Total conductance for B--N and Ga--As. Dash-dotted lines mark the published values from the task-specific \BAsGaN{} potential in the SUS$^2$-SemiCID-TS family of $1.060$ and $0.801$\,GW\,m$^{-2}$\,K$^{-1}$, respectively~\cite{Hu2026SemiCID}. \textbf{d,e}, Corresponding signed spectral conductance, excluding NEP-Oracle-D10. Fitted potentials in this work are \NEP{} students. Circles show each potential's mean over MD runs. Error bars and shading show sample SD across potentials ($n=3$, or $n=15$ for Random-D10). NEP89 is one frozen baseline: diamonds show its three MD simulations, with sample SD across simulations. Simulation windows and spectral normalization are given in Methods.}
    \label{fig:nemd-shc}
\end{figure*}

\begin{figure}[tp]
    \centering
    \includegraphics[width=\columnwidth]{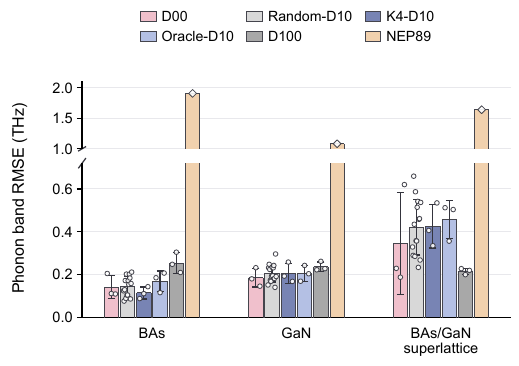}
    \caption{\textbf{Phonon accuracy of student potentials and frozen NEP89.} Frequency-sorted band RMSE against \DFT{} for bulk BAs, GaN and the periodic \BAsGaN{} superlattice. Each potential used its own relaxed geometry. Student bars and error bars show means and sample SD across potentials ($n=3$, or $n=15$ for Random-D10). Open circles show individual potentials. NEP89 bars and diamonds show one frozen checkpoint. The axis break spans $0.72$--$1.00$\,THz and omits no observations. Full dispersions appear in Supplementary Figs.~S20--S25.}
    \label{fig:basgan-phonon-errors}
\end{figure}

\begin{figure*}[tp]
    \centering
    \includegraphics[width=0.8\textwidth]{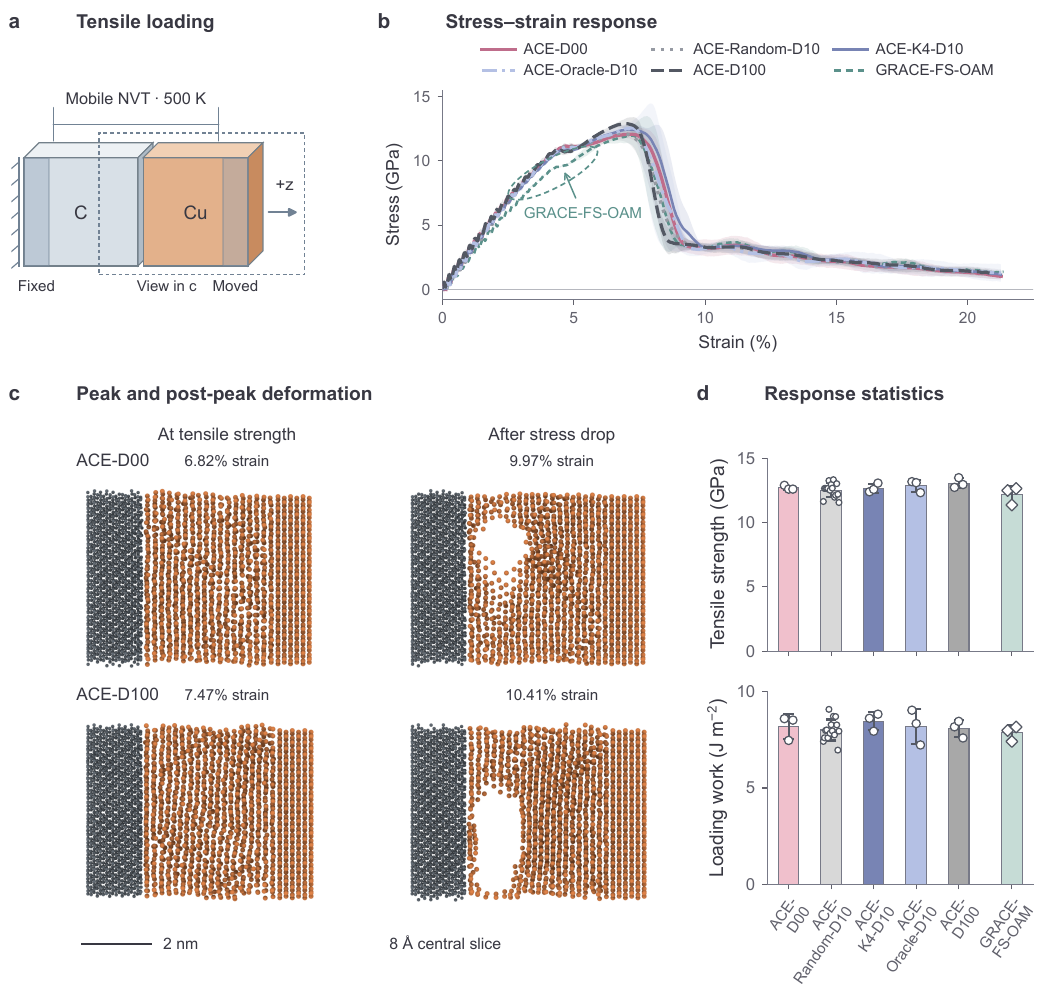}
    \caption{\textbf{\CuDiamond{} tensile response.} \textbf{a}, Cu(100)/diamond(100) specimen and grips. \textbf{b}, Tensile stress versus tensile strain at 500\,K using a 5\,ps moving average. The labeled arrow identifies the lower pre-peak stress of GRACE-FS-OAM relative to the ACE students. For each curve, the strain origin is defined by the near-zero-stress state at the onset of tensile loading. \textbf{c}, ACE-D00 and ACE-D100 slices from one trajectory per strategy, shown near each trajectory's ultimate strain and at its post-drop stage. Tensile strain labels use the original loading reference. \textbf{d}, Tensile strength (maximum tensile stress during loading) and whole-specimen loading work, both evaluated over the full recorded loading histories. Circles show ACE potential means over MD runs. Error bars and shading show sample SD across potentials ($n=3$, or $n=15$ for Random-D10). GRACE-FS-OAM uses one frozen checkpoint: diamonds show three MD simulations, with sample SD across simulations. Snapshot selection and loading-work definitions are given in Methods.}
    \label{fig:cuc-tensile}
\end{figure*}

\begin{figure}[tp]
    \centering
    \includegraphics[width=\columnwidth]{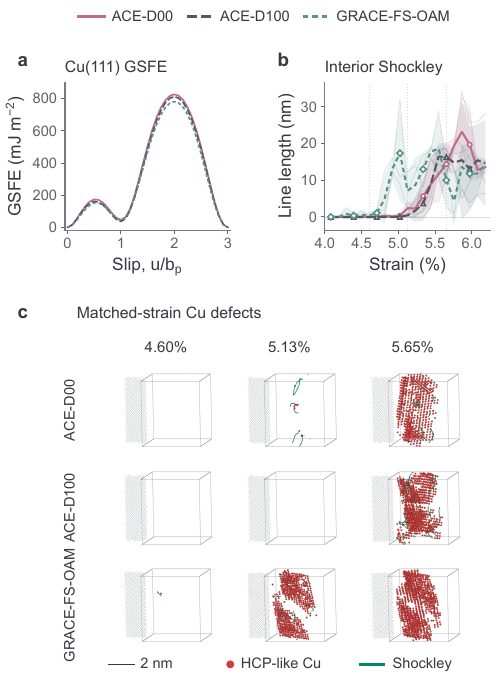}
    \caption{\textbf{\CuDiamond{} slip energetics and early plasticity.} \textbf{a}, Cu(111) generalized stacking-fault energy along $\langle112\rangle$. \textbf{b}, Interior Shockley partial-dislocation length at 500\,K. \textbf{c}, Matched-strain Cu defects (HCP-like Cu in red and Shockley partials in green). Scale bar, 2\,nm. Curves and shading show means and sample SD across three ACE potentials or three GRACE MD runs in \textbf{b}. GRACE uses one checkpoint.}
    \label{fig:cuc-slip-plasticity}
\end{figure}

\begin{figure*}[tp]
    \centering
    \includegraphics[width=0.8\textwidth]{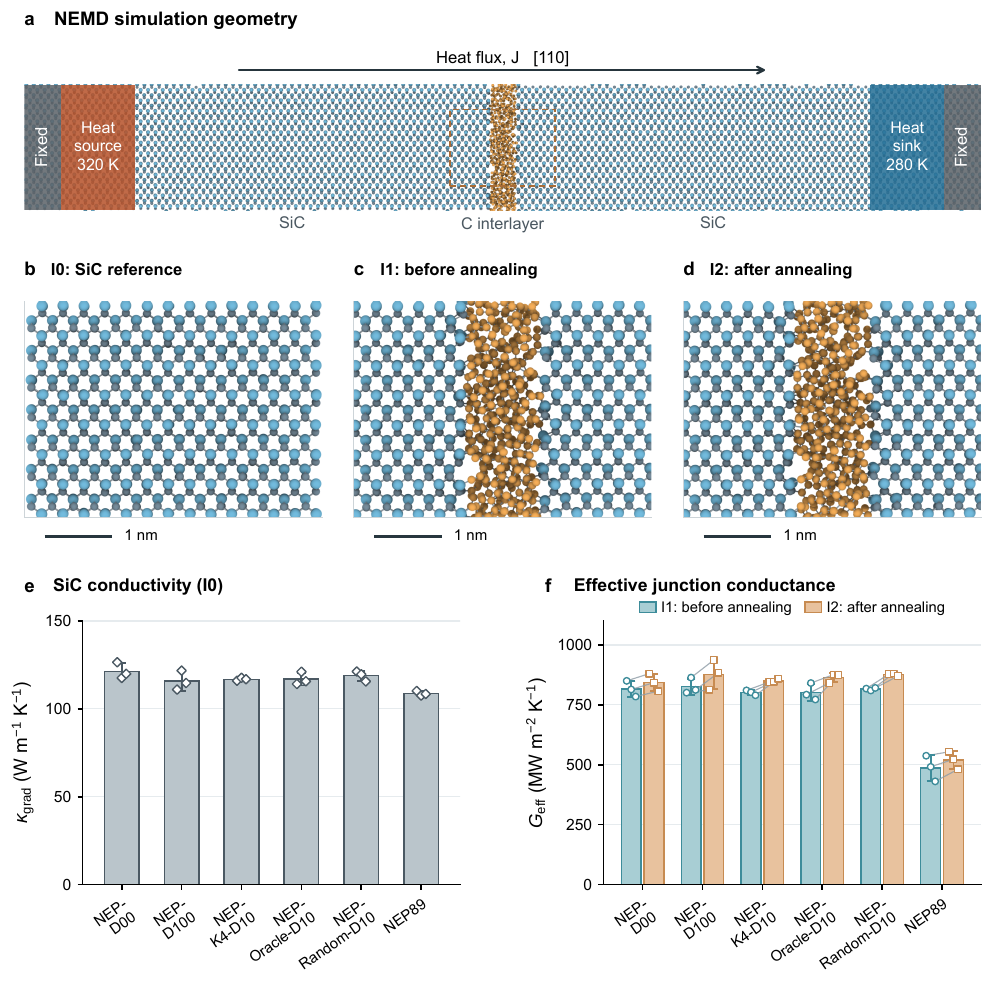}
    \caption{\textbf{Thermal transport through SiC/a-C/SiC interfaces.} \textbf{a}, NEMD geometry. \textbf{b--d}, Pristine SiC (I0) and carbon junctions before (I1) and after (I2) annealing. Si is blue, host carbon gray-blue and interlayer carbon orange. \textbf{e}, Finite-specimen I0 conductivity. \textbf{f}, Effective I1/I2 junction conductance. Bars show means and error bars sample SD ($n=3$ per strategy). Open symbols show individual potential results, means over selections within a Random-D10 training repeat, or preparation and NEMD repeats for the single frozen NEP89 checkpoint. Lines connect paired I1/I2 values. All fitted potentials are \NEP{} students. Preparation and transport definitions are given in Methods.}
    \label{fig:sic-carbon}
\end{figure*}

\begin{figure}[tp]
    \centering
    \includegraphics[width=\columnwidth]{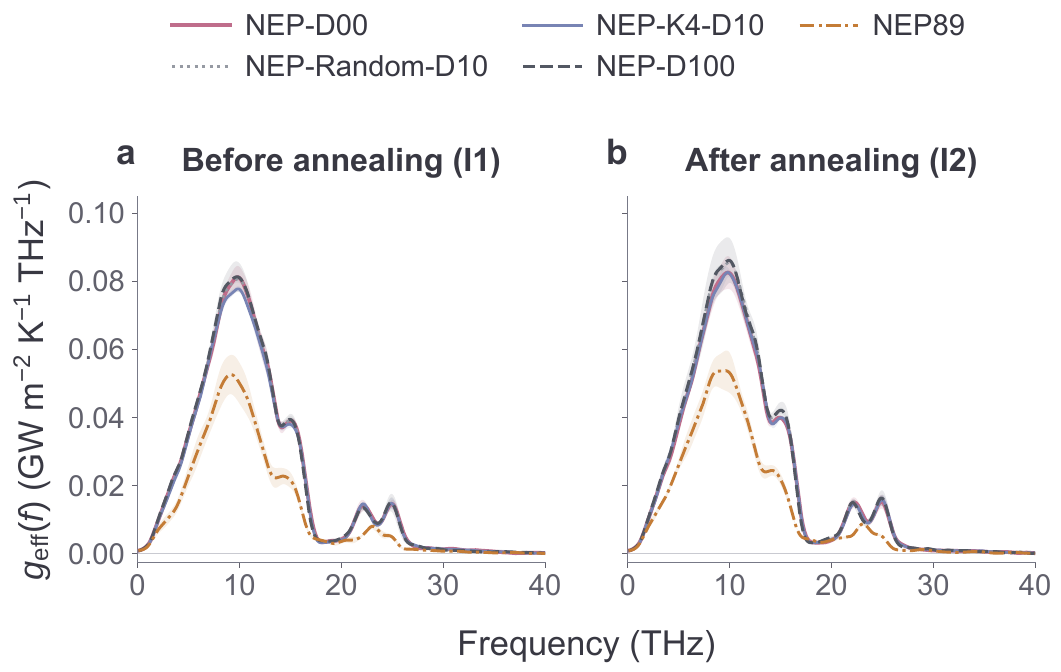}
    \caption{\textbf{SiC/a-C/SiC spectral thermal conductance.} \textbf{a,b}, Signed junction spectra before (I1) and after (I2) annealing. Lines and shading show means and pointwise sample SD across three training-repeat summaries (students) or preparation/MD repeats (NEP89).}
    \label{fig:sic-carbon-shc}
\end{figure}

Selective \DFT{} correction improved student fitting in high-error training regions. We ranked each training set by the largest atomic teacher force-vector error and evaluated every student on identical nested subsets (Methods). On the highest-error $5\%$ of Cu--C configurations, ACE-K4-D10 reduced force-component RMSE against \DFT{} by $16.9\%$ relative to ACE-Random-D10. The corresponding NEP-K4-D10 reductions were $10.7\%$ for \BAsGaN{} and $11.5\%$ for SiC (Fig.~\ref{fig:training-risk}a--c). With the student fitting conditions and nominal $10\%$ correction budget fixed, these reductions showed the benefit of concentrating reference correction on configurations containing large teacher errors. The K4-D10 advantage decreased as the evaluated subsets expanded to the highest-error $10\%$ and $20\%$, indicating that the benefit was concentrated in configurations with the largest teacher errors.

The corresponding test-set differences were smaller and system dependent. The predefined test sets were reserved before student fitting and shared by all labeling strategies within each system (Fig.~\ref{fig:training-risk}d--f). For Cu--C, ACE-K4-D10 and ACE-Random-D10 gave median force-component RMSEs of $139.49$ and $139.84$\,meV\,\AA$^{-1}$, respectively. The corresponding NEP errors were $41.03$ and $43.38$\,meV\,\AA$^{-1}$ for \BAsGaN{}, and $497.51$ and $499.20$\,meV\,\AA$^{-1}$ for SiC. Relative to the retrospective Oracle-D10 control, K4-D10 gave a lower test RMSE for \BAsGaN{} and slightly higher errors for Cu--C and SiC. Within the $20\%$ of test configurations with the largest teacher errors, K4--Random differences remained small. The remaining $80\%$ showed a slight numerical reversal for Cu--C (Supplementary Fig.~S7 and Supplementary Note~3).

On these same test sets, directly distilled students retained force accuracy close to that of their matched \DFT{}-trained counterparts under the same student architectures and fitting protocols. For Cu--C, ACE-D00 and ACE-D100 gave median force-component RMSEs of $139.86$ and $134.67$\,meV\,\AA$^{-1}$, respectively. The corresponding NEP-D00 and NEP-D100 errors were $44.24$ and $38.77$\,meV\,\AA$^{-1}$ for \BAsGaN{}, and $502.25$ and $501.21$\,meV\,\AA$^{-1}$ for SiC. The relative increases over D100 were therefore $3.9\%$, $14.1\%$ and $0.2\%$ across the three systems.

Both D00 and D100 had lower force RMSEs than frozen NEP89, whose corresponding errors were $445.92$, $260.52$ and $790.56$\,meV\,\AA$^{-1}$. The frozen \EqVthree{} teacher remained more accurate, with errors of $51.82$, $26.06$ and $115.98$\,meV\,\AA$^{-1}$, respectively. The teacher--student gap persisted even with full \DFT{} labeling, showing that label replacement alone did not close this gap under the chosen fitting conditions.

Structural examples from the training datasets illustrated where committee ranking succeeded or failed (Supplementary Figs.~S8--S19 and Note~4). They included selected and missed high-error configurations, together with selected and unselected configurations outside the highest-error decile. Maximum atomic teacher force error and maximum atomic disagreement could also occur at different sites.

The Cu--C and \BAsGaN{} tests also informed development, and the SiC test set shared source lineages with training (Methods and Supplementary Note~3). Selective correction therefore showed its clearest benefit in high-error training regions, with smaller and system-dependent test-set differences. Because force errors alone do not establish agreement in physical responses, the following sections compare labeling strategies against matched \DFT{}-trained students across ordered interfaces, nonlinear loading and thermally processed disordered junctions.

\vspace{0.3cm}
\noindent\textbf{Termination-dependent thermal transport at \BAsGaN{} interfaces}
\par\nobreak

\noindent We first tested whether direct distillation preserved termination-dependent heat transport at an ordered \BAsGaN{} interface. B--N and Ga--As terminations alter the local interfacial bonding while retaining the adjoining bulk materials, as examined in previous work on interfacial structure and vibrational coupling~\cite{Wu2024GaNBAs,Yu2026GaNBAs}. Both terminations were evaluated under identical nonequilibrium MD (NEMD) protocols (Fig.~\ref{fig:nemd-shc}a).

NEP-D00 reproduced the higher conductance of B--N relative to Ga--As predicted by NEP-D100. The D00 and D100 conductances were $1.231\pm0.058$ and $1.226\pm0.039$\,GW\,m$^{-2}$\,K$^{-1}$ for B--N, and $0.892\pm0.033$ and $0.955\pm0.057$\,GW\,m$^{-2}$\,K$^{-1}$ for Ga--As, respectively (Fig.~\ref{fig:nemd-shc}b,c). Values are means $\pm$ sample SD across three independently fitted potentials, each averaged over three MD runs. Therefore, although the force RMSE of NEP-D00 was 14.1\% higher (see Fig.~\ref{fig:training-risk}e), its mean conductances differed from those of NEP-D100 by at most 6.6\%, a difference comparable to the sample SDs.

The total-conductance agreement was accompanied by similar frequency-resolved transport. For both terminations, spectral interfacial thermal conductance was concentrated for phonon frequency below $10$\,THz, with NEP-D00 and NEP-D100 sharing a principal peak near $5$\,THz and a shoulder at $7$--$8$\,THz (Fig.~\ref{fig:nemd-shc}d,e). The stronger principal peak for B--N accompanied its higher total conductance. The two student spectra largely overlapped for this termination, whereas NEP-D00 gave a slightly lower peak for Ga--As. At the same nominal \DFT{} fraction, NEP-K4-D10 and NEP-Random-D10 also produced similar total conductances and spectral profiles. Committee-guided correction therefore did not produce a clear separation from random correction in these transport observables.

Equations of state and phonon dispersions provided complementary checks of the students' equilibrium and vibrational responses (Supplementary Figs.~S20--S24 and Supplementary Note~5). Student equations of state closely followed \DFT{} near equilibrium. Across the five labeling strategies, mean full-band RMSEs against \DFT{} were $0.115$--$0.457$\,THz for bulk BAs, GaN and the periodic \BAsGaN{} superlattice (Fig.~\ref{fig:basgan-phonon-errors}). This range covers the means for each labeling strategy and structure. Individual-potential values and the corresponding means $\pm$ sample SD are shown in the figure. Differences among strategies and training repeats were concentrated mainly in higher-frequency optical branches. For modes with \DFT{} frequencies below $10$\,THz, mean errors were $0.052$--$0.105$\,THz across the three structures. The strongest vibrational agreement therefore occurred in the frequency range carrying most of the spectral conductance (Fig.~\ref{fig:nemd-shc}d,e).

Frozen NEP89 tested the practical alternative of directly deploying a pretrained potential. Across three MD simulations per termination, its conductances were $1.212\pm0.057$ and $0.484\pm0.018$\,GW\,m$^{-2}$\,K$^{-1}$ for B--N and Ga--As, respectively. The B--N total remained close to the student values, whereas the Ga--As total was $49.3\%$ below NEP-D100. Its spectral maxima shifted to approximately $4.6$ and $4.2$\,THz, with a weaker $7$--$8$\,THz shoulder. For B--N, enhanced $2$--$4$\,THz and reduced $6$--$8$\,THz contributions accompanied the similar total conductance. The Ga--As principal peak was approximately half the NEP-D100 peak height. Thus, agreement in the B--N total coexisted with a different spectral distribution.

NEP89 also had larger phonon band RMSEs against \DFT{}: $1.909$, $1.082$ and $1.642$\,THz for BAs, GaN and the periodic superlattice, respectively (Fig.~\ref{fig:basgan-phonon-errors} and Supplementary Fig.~S25). These differences included softened BAs optical branches and an upward shift of the highest superlattice branches, with errors of $0.461$--$0.613$\,THz also present below $10$\,THz. Its BAs equation-of-state minimum was displaced toward smaller volumes (Supplementary Fig.~S25a and Supplementary Note~5). These results documented vibrational differences alongside the transport discrepancies, without isolating the origin of the Ga--As conductance reduction.

For comparison, the published task-specific \BAsGaN{} potential from the SUS$^2$-SemiCID-TS family, developed using the SUS$^2$ framework~\cite{Hu2025SUS2,Hu2026SemiCID}, gave reported conductances of $1.060$ and $0.801$\,GW\,m$^{-2}$\,K$^{-1}$ for B--N and Ga--As, respectively. These values provided an external reference with the same qualitative termination ordering. Together, the D00--D100 conductance, spectral and low-frequency phonon comparisons supported retention of the principal interfacial heat-transport response through direct distillation.

\vspace{0.3cm}
\noindent\textbf{Nonlinear tensile response of \CuDiamond{} specimens}

\noindent Nonlinear loading tested whether the agreement between students persisted as deformation changed the sampled atomic environments. ACE-D00 and ACE-D100 simulations started from the same Cu(100)/diamond(100) specimen and followed an identical tensile protocol at 500\,K for 200\,ps, reaching $22.14\%$ tensile strain (Fig.~\ref{fig:cuc-tensile}a). Tensile strength and loading work characterized the whole-specimen response: the former was the maximum tensile stress during loading, while the latter integrated tensile stress over grip displacement across the recorded loading history.

Both students showed a rising stress response followed by a sharp drop, with tensile strengths of $12.70\pm0.17$ and $13.06\pm0.39$\,GPa for ACE-D00 and ACE-D100, respectively, differing by $2.8\%$ (Fig.~\ref{fig:cuc-tensile}b,d). Their loading works were $8.19\pm0.63$ and $8.07\pm0.44$\,J\,m$^{-2}$, a $1.5\%$ difference. Means and sample SDs were calculated across three potentials after averaging three MD repeats within each potential.

Differences at late stages were assessed alongside MD-repeat variability. Over the final 20\,ps, ACE-D00 and ACE-D100 gave mean tensile stresses of $1.23\pm0.29$ and $1.41\pm0.26$\,GPa, respectively (means $\pm$ sample SD across potentials after averaging MD repeats). The difference from the unrounded means was $0.17$\,GPa ($12.3\%$ relative to ACE-D100). For each fixed ACE-D100 potential, the sample SD across three MD velocity seeds ranged from $0.16$ to $0.56$\,GPa. The late-stage difference was therefore comparable in magnitude to the observed MD-repeat variation. Ultimate strain, defined as the tensile strain at tensile strength, also varied among MD repeats. Snapshots near each trajectory's ultimate strain and at its post-drop stage showed localized structural rearrangements (Fig.~\ref{fig:cuc-tensile}c).

The D10 controls tested whether selective correction brought the global loading metrics closer to ACE-D100. ACE-K4-D10 and ACE-Random-D10 gave tensile strengths of $12.68\pm0.35$ and $12.52\pm0.54$\,GPa, and loading works of $8.46\pm0.46$ and $8.01\pm0.54$\,J\,m$^{-2}$, respectively. Means and sample SDs used three K4-D10 potentials and fifteen Random-D10 potentials (five selections with three fits each), after averaging three MD runs per potential. Their mean tensile strengths differed from ACE-D100 by $2.9\%$ and $4.1\%$, and their mean loading works by $4.8\%$ and $0.8\%$, respectively. K4-D10 was numerically closer to ACE-D100 in tensile strength, whereas Random-D10 was closer in loading work.

Early plasticity in the Cu interior provided a microscopic test of the similar global responses. Static Cu(111) generalized stacking-fault energies (GSFEs) along the leading-partial slip path complemented dislocation tracking during loading. ACE-D00 and ACE-D100 gave leading-partial barriers of $174.4\pm1.6$ and $162.2\pm0.5$\,mJ\,m$^{-2}$, respectively (Fig.~\ref{fig:cuc-slip-plasticity}a). Both students showed broadly similar early growth of the Shockley partial-dislocation line length (Fig.~\ref{fig:cuc-slip-plasticity}b). Matched-strain snapshots showed more pronounced localized slip in both ACE examples at $5.65\%$ tensile strain (Fig.~\ref{fig:cuc-slip-plasticity}c).

The frozen pretrained GRACE-FS-OAM checkpoint predicted lower pre-peak stresses and earlier Cu plasticity than the ACE students (Figs.~\ref{fig:cuc-tensile}b and~\ref{fig:cuc-slip-plasticity}b,c). Under the same specimen and loading protocol, three MD simulations gave a tensile strength of $12.17\pm0.70$\,GPa and loading work of $7.85\pm0.39$\,J\,m$^{-2}$. Its lower static leading-partial GSFE barrier of $152.4$\,mJ\,m$^{-2}$, evaluated for that single checkpoint, was consistent with appreciable dislocation activity at lower strains and visible localized slip at $5.13\%$ tensile strain (Fig.~\ref{fig:cuc-slip-plasticity}). Agreement in tensile strength and loading work therefore provided an incomplete assessment of this pretrained potential's mechanical response. The ACE-D00--ACE-D100 comparisons supported retention of the whole-specimen loading response and broadly similar early plasticity through direct distillation under the prescribed protocol.

\vspace{0.3cm}
\noindent{\textbf{Thermal transport through processed SiC/a-C/SiC junctions}}

\noindent The final case examined joint structure preparation and transport prediction in thermally processed disordered junctions. Carbon-rich interlayers have been observed in bonded SiC structures, where their thickness, composition and contacts influence thermal resistance~\cite{Zhao2024SiCAmorphousTransport,Zhao2025SiCInterfaceEvolution}. Each \NEP{} student independently generated its carbon network using the same melt-quench protocol. We compared pristine SiC (I0), the as-prepared SiC/a-C/SiC junction (I1), and its counterpart annealed at $1000$\,K and cooled back to $300$\,K (I2, Fig.~\ref{fig:sic-carbon}a--d), under common NEMD conditions using $320/280$\,K heat reservoirs. I1 and I2 formed paired processing histories within each potential, while comparisons across potentials included differences in the prepared structures.

For the 106,080-atom pristine SiC reference, NEP-D00 and NEP-D100 gave finite-specimen conductivities of $121.3\pm4.6$ and $115.8\pm5.5$\,W\,m$^{-1}$\,K$^{-1}$, respectively (Fig.~\ref{fig:sic-carbon}e). In the 106,941-atom junctions, the carbon layer and its two SiC contacts accounted for about $64\%$ of the bath-to-bath temperature drop under both strategies, with a range of $61$--$65\%$. Thus, both students located the principal thermal resistance in the carbon layer and its contacts while retaining similar crystalline-reference conductivities.

NEP-D00 gave effective junction conductances of $816.3\pm32.9$ and $843.5\pm36.6$\,MW\,m$^{-2}$\,K$^{-1}$ before and after annealing, respectively, compared with $825.7\pm33.4$ and $877.7\pm62.6$\,MW\,m$^{-2}$\,K$^{-1}$ for NEP-D100 (Fig.~\ref{fig:sic-carbon}f). These conductances included the carbon layer and both SiC/carbon contacts. Means and sample SDs used three potentials per strategy, each with its own prepared junction. The D00--D100 differences in the means were smaller than the reported sample SDs. 

The signed spectral conductances provide a complementary test of whether agreement in effective junction conductance extends to frequency-resolved heat transport. For both the as-prepared (I1) and annealed (I2) junctions, NEP-D00 and NEP-D100 showed similar spectral profiles across the principal conducting band (Fig.~\ref{fig:sic-carbon-shc}). Thus, the agreement between directly distilled and \DFT{}-trained students was not limited to the effective conductance, but also extended to its dominant frequency-resolved contributions. Because the spectra include the amorphous carbon layer and both SiC/carbon contacts, they do not independently identify the microscopic origin of the annealing response. Moreover, each potential generated its own disordered structure, so differences in the spectra reflect both structural preparation and transport predictions.

The annealing response was evaluated separately through the paired conductance change within each I1/I2 junction. All three potentials in each strategy predicted an increase, with paired means of $27.2\pm3.9$ and $52.1\pm35.1$\,MW\,m$^{-2}$\,K$^{-1}$ for NEP-D00 and NEP-D100, respectively. Direct distillation therefore retained the direction of the annealing response, while the magnitude and variability of the increments differed.

The D10 controls further distinguished conductance levels from annealing increments (Fig.~\ref{fig:sic-carbon}f). After annealing, NEP-K4-D10 and NEP-Random-D10 gave $850.2\pm7.5$ and $876.5\pm5.5$\,MW\,m$^{-2}$\,K$^{-1}$, respectively, with paired increases of $49.3\pm17.5$ and $60.4\pm10.6$\,MW\,m$^{-2}$\,K$^{-1}$. Random-D10 summaries averaged five selections within each of three training repeats (Methods). The Random-D10 mean was closer to NEP-D100 in post-annealing conductance, whereas the K4-D10 mean was closer in the paired increment. Committee-guided correction thus did not give uniformly closer mean values across these two observables.

Frozen NEP89 tested direct deployment when the pretrained potential also generated the disordered structures. It predicted an increase from $486.5\pm53.7$ to $519.2\pm36.7$\,MW\,m$^{-2}$\,K$^{-1}$, but both conductances remained markedly below the matched student values. Its lower spectral amplitudes over the principal conducting band (Fig.~\ref{fig:sic-carbon-shc}) were consistent with its lower total conductance. The shared direction of the annealing response therefore coexisted with different absolute conductance levels. Within this combined test of structure preparation and transport, NEP-D00 retained the principal thermal bottleneck, similar conductance levels and the direction of the annealing response, with differences in the magnitude and variability of the paired increments.

\vspace{0.5cm}

\noindent\textbf{DISCUSSION}

\noindent Direct distillation combined efficient production MD with retention of the principal interfacial responses of matched \DFT{}-trained students. Fixed training configurations, student architectures and fitting protocols isolated the effect of label source within each system. In the Cu NVE benchmark, ACE and \NEP{} implementations achieved gains of several orders of magnitude in both throughput and the largest completed benchmark size relative to the frozen \EqVthree{} teacher on the same GPU (Fig.~\ref{fig:cu-nve-throughput}). Together, the computational and interfacial-response benchmarks extend earlier physical validations of foundation-model distillation~\cite{Li2025LightPFP,lysogorskiy2026grace}. Fast GRACE-FS-OAM and NEP89 implementations also enable efficient direct deployment, making physical fidelity central to the practical value of constructing a system-specific student.

Physical validation therefore needs to examine both macroscopic observables and the atomic processes contributing to them. In BAs/GaN, similar low-frequency phonon dispersions and dominant spectral-conductance features provided complementary support for the total-conductance comparison. In Cu-diamond, tensile strength and loading work were retained together with broadly similar early plasticity in the Cu interior, although leading-partial slip barriers and late-stage stress responses differed. The observed Shockley partials developed within the Cu interior, so these tensile tests assess whole-specimen loading rather than independently establishing interfacial decohesion. For SiC/a-C/SiC, the directly distilled students reproduced the principal thermal-transport responses of the matched \DFT{}-trained students: the carbon interlayer and its contacts remained the dominant thermal bottleneck, the effective junction conductance increased after annealing, and the main spectral-conductance features were retained in both the as-prepared and annealed states (Fig.~\ref{fig:sic-carbon-shc}). These results show that direct distillation preserved the principal thermal response of the disordered junctions, including the annealing trend and dominant frequency-resolved transport characteristics.

Selective DFT correction revealed a further distinction between identifying inaccurate teacher labels and improving downstream properties. With a 10\% configuration-selection budget, the four-model committee recovered 42.1--55.8\% of the highest-error teacher configurations across the three target datasets. This selection improved student force fitting most clearly in high-error training regions, but the gains on common test configurations were smaller and did not translate into a consistent advantage in thermal or mechanical response. The disagreement score ranks the maximum atomic force deviation and does not directly measure the sensitivity of a physical observable to errors in a particular configuration. Residual student-fitting limitations and differences between the training configurations and the environments sampled during production dynamics may also influence the benefit of label correction. The present results do not separate these contributions, but indicate that improving teacher-label accuracy alone is insufficient to guarantee improved predictions of the investigated properties.

These findings support a property-centered approach to foundation-model distillation. Compact students can first be constructed from the predictions of a frozen teacher and then assessed using physical benchmarks that probe the intended transport or deformation processes. Where deficiencies remain, targeted DFT calculations could be directed toward the relevant structural environments rather than selected solely by force disagreement. Extending this approach to joint thermal and mechanical screening of new interfaces will require independent property benchmarks and explicit accounting of data-generation and simulation costs. Within these boundaries, direct distillation provides a practical route for reusing atomistic foundation models in repeated large-scale interfacial simulations.

\vspace{0.5cm}
\noindent\textbf{METHODS}

\noindent\textbf{Datasets, reference calculations and teacher selection.}

\noindent To assess pretrained atomistic foundation models in semiconductor and interface environments, we constructed a \DFT{}-labeled benchmark from SemiCID~\cite{Hu2026SemiCID}. We randomly sampled 2,000 structures from the local 72,368-configuration collection to form Pilot-2000 for teacher selection and committee development. The model-survey literature cutoff and registry freeze date were both 31 August 2026. We evaluated 128 released checkpoints from 22 model families, using individual checkpoints as the selection unit. Model-family and version sources are summarized in Supplementary Table~S1~\cite{liao2026equiformerv3,Batatia2025-MACE-JCP,lysogorskiy2026grace,zhang2026dpa3,kim2026sevennetomni}. We compared the predicted atomic forces with the \DFT{} reference forces. Force-component RMSE was computed by pooling squared errors over all atoms and Cartesian components. Pilot force errors are shown in Fig.~\ref{fig:pilot-families}a, and the twenty highest-ranked checkpoints and their target-domain errors in Supplementary Table~S2.

To assess direct distillation for interfacial mechanics and heat transport, we used Cu--C and \BAsGaN{} datasets. The Cu--C dataset came directly from our previous study of a transferable copper--carbon potential~\cite{Zhou2026NEPCuCPreprint} and contained 950 training and 50 test configurations. The BAs, GaN and \BAsGaN{} data were taken directly from SemiCID~\cite{Hu2026SemiCID} and contained 2,848 training and 170 test configurations. These datasets were used to assess teacher predictions and train ACE students for \CuDiamond{} tensile loading and \NEP{} students for \BAsGaN{} nonequilibrium molecular dynamics (NEMD) simulations.

The exploratory SiC/carbon dataset drew on SiC-TGAP configurations~\cite{Hamedani2025SiCTGAPData} and the Si/C subset of the carbon-film-growth dataset~\cite{Liu2025CarbonFilmGrowth}. Additional configurations were generated through MD sampling with MACE and NEP89~\cite{Batatia2025-MACE-JCP,liang2026nep89}. We combined 1,300 configurations selected from the existing data with 700 additional structures. These additions covered displaced crystals, amorphous carbon and SiC(110)/amorphous-carbon contacts. Farthest-point sampling used configuration vectors built from atomic NEP89 descriptors. Filtering and random-frame splitting yielded 1,752 training and 195 test configurations. The test set shared source lineages with training. Sampling conditions and dataset preparation are described in Supplementary Note~6.

\DFT{} energies and atomic forces provided a common reference for assessing foundation-model predictions, replacing selected training labels and evaluating the fitted students. SiC/carbon calculations used VASP~\cite{VASP} with the PBE functional~\cite{pbe} and PAW method~\cite{paw}, a 600\,eV plane-wave cutoff, a $k$-point spacing of $0.2$\,\AA$^{-1}$ and an electronic convergence criterion of $10^{-6}$\,eV. The SiC/carbon \texttt{INCAR} is provided in Supplementary Note~6.

Teacher labels were generated with the released EquiformerV3 \texttt{omat24\_gradient.pt} checkpoint, registered locally as \texttt{eqv3\_omat24\_gradient\_local}~\cite{liao2026equiformerv3}. The model used seven Transformer blocks, 128 feature channels and eight attention heads, with maximum spherical-harmonic degree $L_{\max}=4$. Neighbor graphs used a $6$\,\AA{} cutoff and at most 300 neighbors per atom. We also evaluated the same 128 checkpoints on the Cu--C and \BAsGaN{} training sets to audit target-domain force accuracy. The \EqVthree{} teacher selected on Pilot-2000 was retained. Target-domain accuracy and offline inference throughput are compared in Supplementary Fig.~S1a,b. Timing used 100 Cu--C and 200 \BAsGaN{} configurations after warm-up, excluding model loading; results are reported in Supplementary Note~3. Although slower than many surveyed checkpoints, \EqVthree{} served as an offline teacher, so force accuracy took precedence over inference throughput. The checkpoint remained fixed without target-specific fine-tuning while generating energy and force labels for all three target systems. Compact students performed production MD.

\vspace{0.5cm}
\noindent\textbf{Committee construction and uncertainty validation.}

\noindent To prioritize \EqVthree{} labels for \DFT{} correction, we compared atomic-force predictions from a committee of pretrained models~\cite{Liu2026HeterogeneousEnsemble}. We used the \DFT{}-labeled Pilot set to test whether their disagreement identified configurations with large teacher force errors.

For a committee of $K$ selected members, we assigned normalized inverse-RMSE weights
\begin{equation}
w_k=\frac{e_k^{-1}}{\sum_{\ell=1}^{K}e_\ell^{-1}},
\qquad \sum_{k=1}^{K}w_k=1.
\label{eq:committee-weights}
\end{equation}
Here, $e_k$ is model $k$'s force-component RMSE against \DFT{}, pooled over all atoms and Cartesian components of the reference configurations. Members with lower reference force errors therefore received larger weights.

To score configuration $i$, let $\mathbf F_{ijk}$ denote the force predicted by model $k$ on atom $j$. We first calculated the arithmetic mean force on each atom, $\bar{\mathbf F}_{ij}=K^{-1}\sum_{k=1}^{K}\mathbf F_{ijk}$. For each model, we then identified the atom at which its prediction differed most from this mean. The configuration-level score combined these squared maximum deviations:
\begin{equation}
U_i=\left[\sum_{k=1}^{K}w_k
\max_j\left\|\mathbf F_{ijk}-\bar{\mathbf F}_{ij}\right\|_2^2
\right]^{1/2}.
\label{eq:committee-score}
\end{equation}
The maximum was taken separately for each model and could occur at different atoms for different members.

We compared two strategies for selecting members from the 128 screened checkpoints. The quality-ranked strategy added checkpoints in order of increasing reference force RMSE. The family-balanced strategy aimed to construct a heterogeneous committee by including different model families, which span a range of network architectures and representations of atomic environments~\cite{Liu2026HeterogeneousEnsemble}. It first selected the lowest-error checkpoint from each family and added these representatives in order of accuracy. When more members were needed, it returned to the same family order and selected the next-best available checkpoint from each family. Full construction rules are given in Supplementary Note~2.

Pilot configurations were grouped by source-file identity and blocks of 1,000 consecutive source-frame indices, giving 73 groups. These groups were kept intact in five-fold cross-validation. In each round, four training folds determined checkpoint RMSEs, rankings, committee membership and weights. The resulting committee then scored the remaining fold. The pretrained model parameters remained fixed throughout. All 2,000 out-of-fold (OOF) scores were pooled for evaluation.

We assessed the pooled OOF scores against the fixed teacher's largest atomic force-vector error within each configuration. The 200 Pilot configurations with the largest teacher errors formed a fixed high-error reference set. Spearman rank correlation $\rho$ measured whether larger scores tended to accompany larger teacher errors. Average precision (AP) summarized how consistently high-error configurations appeared near the top of the ranking. Recall@10\% reported the fraction of those 200 high-error configurations recovered by selecting the 200 highest-scoring configurations, corresponding to a $10\%$ \DFT{}-selection budget. Full metric definitions, formulas and tie-handling rules are given in Supplementary Note~1.

We tested every committee size from $K=2$ to 128 and selected the smallest size retaining at least $95\%$ of both the maximum AP and maximum Recall@10\% along the quality-ranked OOF curves (Fig.~\ref{fig:pilot-families}b,c). This rule gave $K=4$. At this size, the family-balanced strategy achieved similar screening performance while retaining architectural diversity across four model families. Applying the family-balanced construction to the complete Pilot set yielded \EqVthree{}-OMat-G, PET-OAM-XL, EqV2-86M-Mix and eSEN-30M-OMat~\cite{liao2026equiformerv3,bigi2026pet,barrosluque2026omat24,liao2024equiformerv2,fu2025esen} (Supplementary Table~S3). We recomputed member RMSEs and weights on all 2,000 Pilot configurations and froze the membership and weights before applying the committee to the target datasets. Target score--error associations and high-error retrieval are shown in Supplementary Figs.~S4 and S5 (Note~3). Because Pilot also determined committee size, its metrics were treated as development results.

Pointwise $95\%$ intervals for the committee-size curves used 2,000 bootstrap resamples of the source groups, with committee memberships, weights and high-error labels held fixed. The 2.5th and 97.5th percentiles defined the intervals. Sensitivity checks included 100 random four-model committees, equal model weights and alternative score definitions (Supplementary Fig.~S3, Notes~1 and 2).

\vspace{0.5cm}

\noindent\textbf{Student-potential training and evaluation.}

\noindent With the teacher and committee fixed, we fitted atomic cluster expansion (ACE) students for Cu--C using pacemaker~\cite{Drautz2019ACE,Lysogorskiy2021PACE,Bochkarev2022Pacemaker}. \BAsGaN{} and SiC students used neuroevolution potentials (\NEP{})~\cite{Fan2021NEP,song2024-NEP16}, trained with TorchNEP~\cite{wu2026torchnep}. To compare labeling strategies under matched fitting conditions, we kept the ordered training configurations, student architecture and optimization protocol fixed within each system, including the energy/force loss-weight schedule.

The Cu--C ACE model used a two-density Finnis--Sinclair embedding, an expansion through four-body terms and Bessel radial functions with a 5.5\,\AA{} cutoff. The basis contained 4,864 functions and 10,528 trainable parameters. Fitting used BFGS for up to 3,000 iterations, with a relative force weight $\kappa=0.3$ and $L_1$ and $L_2$ coefficient regularization of $10^{-8}$ each.

For \BAsGaN{} and SiC, the \NEP{} students used the NEP4 architecture with radial and angular cutoffs of 6 and 5\,\AA{}, respectively, expansion orders $n_{\max}=(4,4)$, basis orders $n_{\mathrm{bas}}=(8,8)$, angular settings $l_{\max}=(4,2,1)$ and one hidden layer of 80 neurons. Short-range repulsion used the Ziegler--Biersack--Littmark (ZBL) potential~\cite{Ziegler1985} with inner and outer cutoffs of 1 and 2\,\AA{}. Training followed a two-stage AdamW schedule with batches of 64 configurations, weight decay of $10^{-4}$ and a maximum of 10,000 epochs. The first and second stages used initial learning rates of $10^{-2}$ and $10^{-3}$, with energy/force loss weights of $(0.01,1.0)$ and $(1.0,0.5)$, respectively. Early stopping used a patience of 200 epochs based on training loss, and the virial loss weight was zero.

Under these fixed fitting settings, we compared five labeling strategies. \Dzero{} used energy and force labels from the frozen \EqVthree{} teacher, whereas \Dhundred{} used \DFT{} energies and forces throughout. Teacher energies were shifted to the \DFT{} reference used by \Dhundred{} for consistent comparisons across labeling strategies. K4-D10 replaced the highest-$U$ frames with \DFT{} references, Random-D10 selected the same number uniformly at random, and Oracle-D10 retrospectively selected the largest true teacher errors. Oracle required target \DFT{} labels and was not deployable. D10 selected 95 Cu--C, 285 \BAsGaN{} and 176 SiC frames. Each fixed strategy used three independent fits, whereas Random-D10 used five selections with three fits each, giving 15 potentials.

Foundation-model inference and the generation of teacher energy and force labels were performed on NVIDIA RTX PRO 6000 GPUs, each equipped with 96\,GB of device memory. Production interfacial molecular dynamics simulations were carried out on NVIDIA V100 GPUs. The separate throughput benchmark in Fig.~\ref{fig:cu-nve-throughput} used one NVIDIA RTX 4090 D with 24\,GB of device memory for all implementations. Periodic FCC Cu cells shared the same initial coordinates and physical velocities, a 1\,fs NVE timestep and 200 warm-up steps. 

Test configurations were separated from the training sets before student fitting and shared by all labeling strategies within each system. Energy RMSE used per-configuration energy errors normalized by atom count. Force-component RMSE pooled Cartesian components relative to the original \DFT{} forces. Force-RMSE summaries used the median across fits for fixed strategies and the median of within-selection medians for Random-D10.

The common test sets defined above supported comparative evaluation rather than trajectory-independent generalization. Retrospective analyses of the training sets used 902 Cu--C structures retained after the fitter's internal validation split, all 2,848 \BAsGaN{} structures and all 1,752 SiC structures. Frames were ranked by maximum atomic teacher error, with configuration ID breaking ties. Every student was evaluated on identical nested top-$5\%$, top-$10\%$ and top-$20\%$ subsets, retaining $\lceil qN/100\rceil$ frames for $q=5$, 10 and 20 from each $N$-frame training set.

The frozen pretrained \texttt{nep89\_20250409} checkpoint~\cite{liang2026nep89} was evaluated without training or fine-tuning on the identical predefined training subsets and common test configurations. It was also used unchanged for the transport baselines. In Fig.~\ref{fig:training-risk}, lower axis bands start at zero and retain all student results and \EqVthree{} references. Upper bands show NEP89, and the omitted ranges contain no observations.

\vspace{0.5cm}
\noindent\textbf{\BAsGaN{} thermal-transport simulations.}

\noindent We compared heat transport across 79,920-atom \BAsGaN{} interfaces with B--N and Ga--As terminations using \GPUMD{}~\cite{Xu2025-GPUMD}, fixed end regions and a 1\,fs timestep. After 0.2\,ns of Nos\'e--Hoover-chain NVT equilibration at 300\,K, Langevin reservoirs at 320/280\,K established heat flow for 1\,ns before 5\,ns of production. For the single heat-flow path, total conductance was evaluated over the final 2.5\,ns as
\begin{equation}
J=\frac{|\dot Q_{\mathrm h}|+|\dot Q_{\mathrm c}|}{2A},
\qquad G=\frac{J}{\Delta T},
\label{eq:nemd-conductance}
\end{equation}
where $\dot Q_{\mathrm h}$ and $\dot Q_{\mathrm c}$ are the rates of energy exchange with the hot and cold reservoirs, respectively. Here, $A$ is the equilibrated cross-sectional area, and $\Delta T$ is the temperature jump obtained by extrapolating the two bulk temperature fits to the interface. Spectral heat-current (SHC) analysis used the full 5\,ns production interval, with normalization based on each run's central-interface volume and temperature jump. Spectra were retained without smoothing or rescaling to the total conductance. Student results were averaged over three MD runs per potential before calculating means and sample SD across potentials. The frozen NEP89 baseline followed the same protocol, with sample SD across its three MD runs per termination.

\vspace{0.5cm}
\noindent\textbf{\CuDiamond{} tensile response and plasticity.}

\noindent A 29,792-atom Cu(100)/diamond(100) specimen was loaded normal to the interface in LAMMPS~\cite{LAMMPS} at 500\,K with a 0.5\,fs timestep. The cell dimensions were fixed, with periodic transverse boundaries and a nonperiodic loading direction $z$. Following NVT equilibration, the lower grip remained fixed while the upper grip moved along $+z$ according to a prescribed displacement. Tensile strain and grip displacement were defined as
\begin{equation}
\varepsilon(t)=\frac{L(t)-L_0}{L_0}=\frac{\Delta z(t)}{L_0}, \Delta z(t)=L_0[\exp(rt)-1].
\label{eq:tensile-loading}
\end{equation}

Here, $L(t)$ is the grip-center separation, $L_0=L(0)$ and $r=0.001$\,ps$^{-1}$ is the exponential loading parameter. Thus, $\varepsilon(t)=\exp(rt)-1$, reaching $0.2214$ after 200\,ps. The tensile strain rate increases as $r\exp(rt)$. During loading, mobile atoms were thermostatted through their transverse velocity components. Tensile stress and signed loading work per initial area were calculated as
\begin{equation}
\sigma_{\mathrm{t}}(t)=-\frac{F_z^{\mathrm{grip}}(t)}{A_0}, \qquad W_{\mathrm{load}}=\int_{\Delta z_{\mathrm{first}}}^{\Delta z_{\mathrm{last}}}
\sigma_{\mathrm{t}}\,\mathrm d(\Delta z),
\label{eq:tensile-observables}
\end{equation}
where $F_z^{\mathrm{grip}}$ is the specimen's force on the moving grip and $A_0=L_xL_y$ is the initial cross-sectional area, with tension positive. The integral spans the full recorded displacement interval and measures whole-specimen loading work. Tensile stress curves used a 5\,ps valid moving average, with tensile strain coordinates averaged over the same window. Tensile strength was defined as the maximum of the averaged stress history of each trajectory. Ultimate strain was the corresponding tensile strain, measured from the original loading reference. Work used the recorded stress and displacement without additional smoothing. Student results were averaged first over MD repeats within each potential and then across potentials, following the \BAsGaN{} procedure. Frozen GRACE-FS-OAM used the same specimen and loading protocol, with mean and sample SD across three MD runs.

To characterize slip energetics and early plasticity, we evaluated the Cu(111) generalized stacking-fault energy (GSFE) along a straight $\langle112\rangle$ path over one complete lattice translation, $0\le u/b_{\rm p}\le3$. Here, $b_{\rm p}=a/\sqrt{6}$ is the magnitude of an $a/6\langle112\rangle$ Shockley partial Burgers vector. The 60-atom, 15-layer slab contained one internal fault and approximately 20\,\AA{} of vacuum. We sampled 37 displacements at intervals of $b_{\rm p}/12$. Using the common \DFT{} in-plane lattice, we fixed the cell and all in-plane atomic coordinates, held the original four reference-layer atoms fixed, and relaxed the other 56 normal coordinates until the maximum free normal force was no larger than $10^{-5}$\,eV\,\AA$^{-1}$. For the single-fault slab, the GSFE was
\begin{equation}
\gamma(u)=\frac{E(u)-E(0)}{A_{\mathrm{fault}}},
\label{eq:gsfe}
\end{equation}
where $u$ is the imposed slip displacement, $E(u)$ is the relaxed energy and $A_{\mathrm{fault}}=22.8816$\,\AA$^2$ is the fault area. Each curve used its own zero-slip energy. Full-period endpoints at $u/b_{\rm p}=3$ represented an equivalent perfect lattice and were checked for closure. The leading-partial barrier was the maximum sampled GSFE over $0\le u/b_{\rm p}\le1$, and the intrinsic stacking-fault energy was evaluated at $u/b_{\rm p}=1$. Static ACE curves show means and sample SD across three trained potentials per strategy. GRACE-FS-OAM supplied a single static curve. We then applied OVITO's dislocation extraction algorithm~\cite{Ovito,Stukowski2012DXA} directly to stored Cu coordinates, using an FCC reference and enabling partial dislocations. Shockley line length associated with a fixed Cu-interior region was tracked between $4.08\%$ and $6.18\%$ tensile strains, with averaging following the tensile observables. Thin curves show individual ACE potential means or GRACE MD runs. Polyhedral template matching~\cite{Larsen2016PTM} classified the local environments of all Cu atoms before HCP-like atoms in the interior were selected for the matched-strain defect views. Snapshots at $4.60\%$, $5.13\%$ and $5.65\%$ tensile strain used the same view and scale. Other Cu atoms were hidden. Pale gray Cu outlines and carbon atoms provided geometric context.

\vspace{0.5cm}
\noindent\textbf{SiC/a-C/SiC preparation and thermal transport.}

\noindent To examine processing-dependent disorder, we compared 106,080-atom pristine SiC specimens (I0) with 106,941-atom SiC/a-C/SiC junctions before (I1) and after (I2) annealing. Each potential independently generated its carbon interlayer by melting at 6000\,K and quenching to 300\,K under a common thermal schedule. I0 and I1 then underwent NVT and zero-pressure orthorhombic NPT relaxation at 300\,K. Each potential prepared I2 from its own I1 endpoint by heating to 1000\,K, annealing for 100\,ps, cooling to 300\,K and relaxing at zero pressure. I1 and I2 therefore formed paired processing histories within each potential, while comparisons across potentials included differences in both structure preparation and transport prediction.

Transport simulations used the same 300\,K equilibration, 320/280\,K reservoirs and 1\,ns heat-flow establishment as the \BAsGaN{} benchmark, with a 1\,fs timestep. Heat flux followed Eq.~\eqref{eq:nemd-conductance} using each run's initial cross-sectional area, and transport quantities were evaluated over the full 5\,ns production interval. The finite-specimen conductivity and effective junction conductance were
\begin{equation}
\kappa_{\mathrm{grad}}=\frac{J}{\left|(\mathrm dT/\mathrm dz)_{\mathrm{fit}}\right|},\qquad
G_{\mathrm{eff}}=\frac{J}{\Delta T_{\mathrm{junction}}}.
\label{eq:sic-transport}
\end{equation}
Here, $J>0$ denotes the heat-flux magnitude, and $z$ increases from the hot reservoir toward the cold reservoir. The temperature gradient fitted to the I0 interior is therefore negative, and its magnitude is used to calculate $\kappa_{\mathrm{grad}}$. For I1 and I2, $\Delta T_{\mathrm{junction}}$ was obtained by extrapolating the two SiC temperature fits to their junction reference planes. Thus, $G_{\mathrm{eff}}$ includes the carbon layer and both SiC/carbon contacts. Fixed strategies used three independently trained potentials, each providing one I0 run and one paired I1/I2 preparation and transport sequence. Random-D10 first averaged its five selections within each of three training repeats. Means and sample SD used the resulting three values. Annealing increments were calculated within each I1/I2 pair before aggregation. Frozen NEP89 used three independent preparation and MD sequences, with paired I1/I2 states and sample SD across sequences. Signed I1/I2 SHC spectra used the full 5\,ns production interval and each run's actual central-window volume and $\Delta T_{\mathrm{junction}}$ for normalization. The window enclosed the carbon layer and both contacts. Spectral means and pointwise sample SD used the same replication units as $G_{\mathrm{eff}}$, without additional smoothing or rescaling to the total conductance.

\vspace{0.5cm}
\noindent\textbf{Use of generative AI tools.}

\noindent Generative AI tools were used to assist with language editing and manuscript restructuring. All scientific claims, numerical results, analyses and interpretations were reviewed by the authors, who take responsibility for the final manuscript.

\vspace{0.5cm}
\noindent{\textbf{DATA AVAILABILITY}}

\noindent The D00 and D100 student potentials and their corresponding training datasets for \BAsGaN{}, \CuDiamond{} and SiC/a-C/SiC are publicly available in the Distill2MD-Interfaces repository (\url{https://github.com/zhoufy20/Distill2MD-Interfaces}). Additional data are available from the corresponding author upon reasonable request.

\vspace{0.5cm}
\noindent{\textbf{CODE AVAILABILITY}}

\noindent The atomic cluster expansion (ACE) student potentials were fitted using pacemaker~\cite{Bochkarev2022Pacemaker}, whereas the neuroevolution potential (NEP) student models were trained using TorchNEP~\cite{wu2026torchnep}. Molecular dynamics simulations were performed using \GPUMD{}~\cite{Xu2025-GPUMD} and LAMMPS~\cite{LAMMPS}, with the NEP--LAMMPS interface (\url{https://github.com/brucefan1983/NEP_CPU}) employed for NEP-based simulations in LAMMPS. Selected \GPUMD{} simulations employed the Wisevolve Potential Engine (WPE) backend~\cite{Bu2026WPE}. Phonon calculations were carried out using Phonopy~\cite{Togo2023PhonopyImplementation,Togo2023Phonopy}, while \GPUMD{}kit~\cite{Yan2026GPUMDkit} and \href{https://github.com/bigd4/PyNEP}{PyNEP} were used for data processing and post-processing analyses.

\vspace{0.5cm}
\noindent{\textbf{\large Author contributions}}

\noindent F.Z. and W.Q. contributed equally to this work. F.Z. and W.Q. developed the computational workflow, performed the simulations and analyzed the data. F.S. contributed to methodology development, uncertainty assessment and model validation. Y.L. contributed to scientific discussions and provided feedback on the manuscript. F.S. and P.Y. contributed to study conception and design. S.Y. and P.Y. contributed to physical interpretation, resources, and project supervision. F.Z., W.Q. and P.Y. prepared the initial manuscript. All authors discussed the results and contributed to manuscript revision.

\vspace{0.5cm}
\noindent{\textbf{\large Competing interests}}

\noindent The authors declare no financial or non-financial competing interests.

\vspace{0.5cm}
\noindent{\textbf{\large Acknowledgments}}

\noindent The authors acknowledge financial support from the National Natural Science Foundation of China (Grant No. 12602174), the National Natural Science Foundation of China (Grant No. 12674045), the National Natural Science Foundation of China (GYKP022), and the Xi'an Jiaotong University Young Talent Program (LX6J0240001). W. Qiu acknowledges postdoctoral fellowships from the Sackler Center for Computational Molecular and Materials Science and the Ratner Center for Single Molecule Science at Tel Aviv University. The authors also acknowledge the Open Source Supercomputing Center of S-A-I for providing computing resources.

\vspace{0.5cm}
\noindent{\textbf{\large Correspondence}} and requests for materials should be addressed to Fei Shuang, Shengying Yue, and Penghua Ying.

\vspace{0.5cm}
\noindent{\textbf{\large References}}
\bibliographystyle{naturemag}
\bibliography{ref}
% \nolinenumbers

\end{document}